\documentclass[12pt]{iopart}
\usepackage{cite}
\usepackage{fixmath}
\usepackage{graphicx, rotating}
\usepackage{iopams} 
\usepackage{multirow}
\usepackage{setstack}
\usepackage[dvipsnames]{xcolor}
\usepackage[normalem]{ulem}
\usepackage{footmisc}
\usepackage{etoolbox}
\usepackage[margin=1in]{geometry} 
\usepackage[T1]{fontenc}

\makeatletter
\newcommand{\mainmatter}{%
  \setcounter{footnote}{0}%
    \patchcmd{\@makefntext}{\fnsymbol}{\arabic}{}{}%
  \patchcmd{\@thefnmark}{\fnsymbol}{\arabic}{}{}%
    \def\@makefnmark{\textsuperscript{\arabic{footnote}}}%
}
\makeatother

\newcommand{\dv}[2]{\frac{\mathrm{d}#1}{\mathrm{d}#2}}
\newcommand{\vb}[1]{\mathbold{#1}}
\newcommand{\pdv}[2]{\frac{\partial #1}{\partial #2}}
\newcommand{\Ham}{\mathcal{H}}
\newcommand{\Exp}[1]{\mathrm{e}^{#1}}
\newcommand{\grad}{\boldsymbol{\nabla}} 
\newcommand{\ABS}[1]{{\,{\vrule width 0.4mm \vphantom{\abs{#1}}} #1 \,{\vrule width 0.4mm \vphantom{\abs{#1}}}\,}}

\newcommand{\I}{\mathrm{i}}
\newcommand{\ket}[1]{{\left|{#1}\right\rangle}}

\newcommand{\order}[1]{\mathcal{O} \left( #1 \right)}

\newcommand{\FFT}[1]{\mathcal{F} \left\{ #1 \right\}}

\newcommand{\ds}{\displaystyle}
\newcommand{\sds}[2][\depth]{\raisebox{0pt}[\height][#1]{\small$\ds #2$}}
\newcommand{\K}{\mathrm{K}}
\newcommand{\abs}[1]{ \left| #1 \right| } 
\newcommand{\braket}[2]{ \left \langle {#1} \middle | {#2} \right \rangle }

\newcommand{\dd}[1]{\mathrm{d}{#1}} 

\renewcommand{\submitto}[1]{\vspace{28pt plus 10pt minus 18pt}\noindent{\small\textrm{Posted on the arXiv on #1.}\par}}

\begin{document}
	\title[Fourth-order leapfrog algorithms for numerical time evolution]{Fourth-order leapfrog algorithms for numerical\\ time evolution of classical and quantum systems} 
	\author{Jun Hao Hue$ ^{1,2,\footnotemark} $ \footnotetext{junhao.hue@u.nus.edu}, Ege Eren$ ^{3, \footnotemark} $ \footnotetext{ege.eren@boun.edu.tr}, Shao Hen Chiew$ ^{2, \footnotemark} $ \footnotetext{shaohenc@gmail.com}, Jonathan Wei Zhong Lau$ ^{2, \footnotemark} $ \footnotetext{jonathanlau66@gmail.com}, Leo Chang$ ^{2, \footnotemark} $ \footnotetext{a1997411z@gmail.com}, Thanh Tri Chau$ ^{2,\footnotemark} $\footnotetext{chau.thanhtri@protonmail.com}, Martin-Isbj\"{o}rn Trappe$ ^{2,4,\footnotemark} $\footnotetext{martin.trappe@quantumlah.org} and Berthold-Georg Englert$ ^{2,5,6,\footnotemark} $\footnotetext{cqtebg@nus.edu.sg}}
	
	\address{$ ^1 $ Graduate School for Integrative Sciences \& Engineering, National University of Singapore, 21 Lower Kent Ridge Road, Singapore 119077, Singapore}
	\address{$ ^2 $ Centre for Quantum Technologies, National University of Singapore, 3 Science Drive 2, Singapore 117543, Singapore}
	\address{$ ^3 $ Department of Physics, Bo\u{g}azi\k{c}i University, 34342 Bebek, Istanbul, Turkey}
	\address{$ ^4 $ Department of Biological Sciences, National University of Singapore, 16 Science Drive 4, Singapore 117558, Singapore}
	\address{$ ^5 $ Department of Physics, National University of Singapore, 2 Science Drive 3, Singapore 117542, Singapore}
	\address{$ ^6 $ MajuLab, CNRS-UNS-NUS-NTU International Joint Unit, UMI 3654, Singapore}

	\begin{abstract}		
		Chau~\textit{et~al.}~[New~J.~Phys.~\textbf{20},~073003~(2018)] presented a new and straight\-forward derivation of a fourth-order approximation `$U_7$' of the time-evolution operator and hinted at its potential value as a symplectic integrator. $U_7$ is based on the Suzuki--Trotter split-operator method and leads to an algorithm for numerical time propagation that is superior to established methods. We benchmark the performance of $U_7$ and other algorithms, including a Runge--Kutta method and another recently developed Suzuki--Trotter-based scheme, that are exact up to fourth order in the evolution parameter, against various classical and quantum systems. We find $U_7$ to deliver any given target accuracy with the lowest computational cost, across all systems and algorithms tested here. This study is accompanied by open-source numerical software that we hope will prove valuable in the classroom.\\
	\end{abstract}
	
	\submitto{10th July 2020}	
	\maketitle

	\mainmatter
	
	\section{Introduction}

	Whatever your physical system of interest, at the end of the day you want to predict its time evolution --- and the typical system of practical relevance cannot be solved analytically. It is no surprise, then, that numerical methods for time propagation have been in high demand since the advent of the natural sciences. In fact, the time-evolution algorithms (TEA) presented here are closely related to the so-called leapfrog algorithm (see, for example, \cite{DeVries:94, Garcia:00, Giordano+1:06}), which dates back at least to Newton \cite{Newton:1687}. A TEA propagates a system (for example position and momentum of a classical particle) for time steps $\Delta t$ and aims at approaching the true system state at time $T$. Every TEA can achieve arbitrary accuracy if $\Delta t$ is small enough and accumulated rounding-off errors are of no concern, but the computational cost of the TEA may become prohibitive in practice. Here, we develop and apply an easy-to-implement TEA that reaches a given accuracy more efficiently than established algorithms like the Runge--Kutta method \cite{Runge:1895, DeVries:94, Kutta:1901}.
	
	TEAs can be categorized by the scaling of their error with $\Delta t$. The simplest algorithm in the family of Runge--Kutta methods is the Euler method, whose error is quadratic in $\Delta t$. It is exact up to first order in $\Delta t$ and is therefore termed a 1st-order method. Our work here features a 4th-order TEA (`$U_7$'), as developed in \cite{Chau+3:18} in the context of density functional theory. This algorithm had been presented independently in several publications before \cite{Chin:97,Omelyan+2:02}. Here, we report a new and more direct derivation and show that $U_7$ outperforms popular methods of the same order, like the 4th-order Runge--Kutta method (`RK4'). As a bonus, $U_7$ is a symplectic integrator: In contrast to RK4, it preserves (oriented) volumina of generalized phase space during Hamiltonian evolution. Kepler orbits, for instance, do not decay when propagated with $U_7$.
	
	There is a long history of developing symplectic TEA (see, for example, \cite{Hairer+2:02}), with the Suzuki--Trotter (ST) split operator method \cite{Trotter:59, Suzuki:76, Hatano+1:05} among the most popular. The ST method reveals the well-known link between classical and quantum dynamics (see, for example, \cite{Hatano+1:05,Dattoli+3:97}) and permits using the same TEA for both. The ST approach to time evolution is also instructive in that it shows a straightforward path towards higher-order time-evolution algorithms that are easy to implement and to apply in class-room settings: Our article is accompanied by an open-source program\footnote{https://github.com/huehou/Fourth-Order-Leapfrog } ready to be used, for example, in undergraduate courses of classical or quantum mechanics. We hope that in this way we can contribute to bridging between the often modest efforts in developing numerical skills in institutions of higher education and the demands of today's scientific environment. 
	
	In this article we benchmark the performance of our ST-based $U_7$ algorithm against alternative 4th-order TEA and against the exact solutions for a selection of systems, including textbook examples like the classical pendulum as well as more advanced applications like Rydberg wave packets. In \sref{sec:leapfrog} we set the stage by elucidating the connection between classical and quantum dynamics and develop the various ST approximations that we study subsequently. In \sref{sec:compare} we specify the physical systems considered in our benchmarking exercise and define our benchmarking protocols. We present our performance results for classical and quantum systems in sections~\ref{sec:classical} and \ref{sec:quantum}, respectively.

	\section{Time-evolution algorithms from Suzuki--Trotter factorizations} \label{sec:leapfrog}

	\renewcommand{\thefootnote}{\roman{footnote}}
	
	Consider a classical single-particle system with momentum $\vb{p}$ and a potential energy ${V(\vb r)}$ that only depends on the particle position $\vb{r}$. The Hamiltonian is
	\begin{equation}\label{eq:ClassHam}
		\Ham = \frac{\vb{p}^2}{2m} +  V{\left(\vb{r}\right)} \,,
	\end{equation}
	where $m$ is the particle mass. Then, the Hamilton equations of motion ${ \dd{\vb{r}}/\dd{t} = \partial \Ham / \partial \vb{p} }$ and ${\dd{\vb{p}}/\dd{t} = - \partial \Ham/ \partial \vb{r}}$ comprise the differential equation
	\begin{equation}\label{eq:HamEq}
		\dv{}{t} \left( \begin{array}{c}
			\vb{r} \\
			\vb{p}
			\end{array} \right) = \left( \frac{\vb{p}}{m} \pdv{}{\vb{r}} - \grad V(\vb{r}) \pdv{}{\vb{p}}  \right) \left( \begin{array}{c}
			\vb{r}\\
			\vb{p}
			\end{array} \right)  \equiv - \Ham_{\mathrm{P}} \left( \begin{array}{c}
			\vb{r}\\
			\vb{p}
	\end{array} \right),
	\end{equation}
	where we define\footnote{ \Eref{eq:HamEq} is a special case of the more general Hamilton equation of motion ${ \dv{}{t} A{ \left( \vb{r}. \vb{p}, t \right) } = \pdv{}{t}A - { \left\{ \Ham, A \right\} }_{\mathrm{PB}} = \pdv{A}{t} - \Ham_{\mathrm{P}} A }$, where ${ \left\{ \,,\, \right\}_{\mathrm{PB}} }$ is the Poisson bracket.   }
	\begin{equation}
		f_\mathrm{P} = \pdv{f}{\vb{r}} \pdv{}{\vb{p}} - \pdv{f}{\vb{p}} \pdv{}{\vb{r}}
	\end{equation}
	for a function ${ f(\vb r, \vb p) }$. \Eref{eq:HamEq} is formally solved by 
	\begin{equation}
		\left(\begin{array}{c}
			\vb{r}\\
			\vb{p}
			\end{array}\right) = \Exp{- t  {\Ham_\mathrm{P}} } \left( \begin{array}{c}
			\vb{r}\\
			\vb{p}
	\end{array} \right).
	\label{eq:ClassicalSol}
	\end{equation}

	We recognize the formal equivalence between \eref{eq:ClassicalSol} and the solution to the Schr\"{o}dinger equation, see \cite{Hatano+1:05, Dattoli+3:97}: Consider a single-particle quantum system with the Hamiltonian\footnote{We write $H$ for the Hamilton operator and $\Ham_{\mathrm{S}}$ for its matrix representation in a basis $\{\ket{\vb a,t}\}$. Accordingly, $\Psi(t)$ denotes the collection of amplitudes ${\psi(\vb a,t)=\braket{\vb a,t}{\psi}}$ that make up the wave function of the system state $\ket{\psi}$ at time $t$. }
	\begin{equation}
		H = \frac{\vb{P}^2}{2m} + V{ \left( \vb{R} \right)} \,,
	\end{equation}
	where $\vb{R}$ and $\vb{P}$ are the position and momentum operators, respectively. The Schr\"{o}dinger equation ${\I \hbar \pdv{}{t} \Psi(t) = \Ham_{\mathrm{S}} \Psi(t)}$ is then solved by
	\begin{equation}
		\Psi(t) = \Exp{- \frac{\I t}{\hbar} \Ham_\mathrm{S} } \Psi(0) \equiv \mathcal U(t) \Psi(0),
		\label{eq:QuantumSol}
	\end{equation}
	where $\mathcal U(t)$ is a matrix representation of the time-evolution operator $U(t)$. The structural equivalence between \eref{eq:ClassicalSol} and \eref{eq:QuantumSol} is completed by the identifications 
	\begin{eqnarray}
		\Exp{t \frac{\vb{p}}{m}  \pdv{}{\vb{r}}} {}&\leftrightarrow \hspace{2em} \mathrm{translate}\; \vb{r}\; \mathrm{by}\; t \frac{\vb{p}}{m}  {}&\leftrightarrow \hspace{0.2cm} \Exp{- \frac{\I t}{\hbar} \frac{\vb{P}^2}{2m}  } \label{eq:connection1} \\
		\hspace{-70pt} \mathrm{and} \nonumber\\
		\Exp{-t \grad V(\vb{r}) \pdv{}{\vb{p}}} \hspace{0.2cm} {}&\leftrightarrow \hspace{0.2cm} \mathrm{translate}\; \vb{p}\; \mathrm{by}\; -t \grad V(\vb{r}) \hspace{0.2cm} {}&\leftrightarrow \hspace{0.2cm} \Exp{- \frac{\I t}{\hbar} V(\vb{R})  }. \label{eq:connection2}
	\end{eqnarray}
	Therefore, any approximation of $\mathcal U(t)$ in \eref{eq:QuantumSol} defines an equivalent approximation of $\Exp{- t \Ham_{\mathrm{P}} }$ in \eref{eq:ClassicalSol}, and vice versa.
	
	In the following, we establish a series of increasingly accurate ST approximations
	\begin{equation}
		U_N = \prod_{i=1}^{\lceil N/2 \rceil} \Exp{- \frac{\I t}{\hbar} \alpha_i V{ \left( \vb{R} \right)} } \Exp{- \frac{\I t}{\hbar} \beta_i \frac{\vb{P}^2}{2m}  } \equiv \prod_i \Exp{t\,A_i} \Exp{t\,B_i} = \Exp{t \, A_1} \Exp{t\,B_1} \Exp{t\,A_2} \Exp{t\,B_2} \dots  , 
		\label{eq:STDecomp}
	\end{equation}	
	of $U(t)$, where the coefficients $\alpha_i$ and $\beta_i$ can be chosen to minimize the error of $U_N$ at a specific order in $t$. For example, since $\vb{P}^2$ does not commute with ${V{ \left( \vb{R} \right)}}$,
	\begin{equation}
		U_2 \equiv \Exp{- \frac{\I t}{\hbar} V{ \left( \vb{R} \right)} } \Exp{- \frac{\I t}{\hbar} \frac{\vb{P}^2}{2m}  }
	\end{equation}
	retrieves only the 1st-order of the Taylor expansion of ${U(t)=\Exp{- \frac{\I t}{\hbar}H }}$, and is therefore a 1st-order approximation with errors in the ${\mathcal O(t^2)}$ terms. The factorizations $U_N$ are unitary and reversible, preserving the probability density. \Tref{tab:STCoeff} shows several ST approximations taken from \cite{Chau+3:18} and \cite{Hatano+1:05} up to 4th order.\footnote{Aside from ${U_2}$, all factorizations in \tref{tab:STCoeff} are symmetric and therefore void of even-order errors. Symmetric 3rd-order approximations, for instance, are automatically exact up to ${\order{t^4}}$.} In the context of density-potential functional theory, $U_2$ delivers the particle density in Thomas--Fermi approximation \cite{Chau+3:18}, and a variant of $U_3$ employed in \cite{Trappe+2:19} reveals quantum corrections for two-dimensional materials beyond the Thomas--Fermi approximation. High-quality particle densities for harmonium are calculated with the help of $U_5$ in \cite{Chau+3:18} and $U_7$ in \cite{Trappe+4:?}. To propagate a state at time $t_0$ by one time step $\Delta t$, we apply the exponential factors of $U_N$ in sequence. For example, the quantum algorithm for $U_3$ reads 
	\begin{equation}\label{U3FFT}
		\psi \left( \vb{x}, t_0 + \Delta t \right) = \Exp{\Delta t\,A_2} \mathcal{F}^{-1}\Big\{ \Exp{\Delta t\,B_1} \FFT{ \Exp{\Delta t\,A_1} \psi \left( \vb{x}, t_0 \right) } \Big\} \,,
	\end{equation}
	where $\mathcal{F}$ and $\mathcal{F}^{-1}$ are the Fourier and inverse Fourier transforms that convert wave functions in $\vb{x}$ to wave functions in $\vb{p}$ and back, respectively.

	\begin{table}[htb!]
		\centering
		\caption{A selection of increasingly accurate $n$th order ST approximations of the time-evolution operator, see \eref{eq:STDecomp}, with coefficients $\alpha_i$ and $\beta_i$ chosen such that the error at order ${\mathcal{O}\big( t^{n}\big)}$ vanishes. Note that $U_7$ requires only five factors, which renders it computationally more efficient than $U_{11}$. ($*$) indicates that the exponent ${\alpha_2 V}$ for $U_7$ in \eref{eq:STDecomp} is replaced by \sds{ 2V/3 - \big(t^2/(72m)\big) (\grad V)^2 }, see \fref{fig:Leapfrog}. The parameters for $U_7'$ and $U_{11}$ are \sds{{s = (2-2^{1/3})^{-1} \cong 1.35}} and \sds{ {k = (4 - 4^{1/3})^{-1} \cong 0.41} }, respectively, and therefore the three central coefficients are negative. }
		\label{tab:STCoeff}
		\hspace{2cm}
		$\begin{array}{@{}c c c c c c c c c c c c c}
			\br
			 & \alpha_1 & \beta_1 & \alpha_2 & \beta_2 & \alpha_3 & \beta_3 & \alpha_4 & \beta_4 & \alpha_5 & \beta_5 & \alpha_6 & n \\
			\mr
			U_2 & 1 & 1 & & & & & & & & & & 1 \\
			U_3 & \frac{1}{2} & 1 & \frac{1}{2} & & & & & & & & & 2\\
			U_7' & \frac{s}{2} & s & \frac{1-s}{2} & 1-2s & \frac{1-s}{2}  & s & \frac{s}{2} & & & & & 4\\
			U_{11} & \frac{k}{2} & k & k & k & \frac{1-3k}{2} & 1-4k & \frac{1-3k}{2}  & k & k & k & \frac{k}{2} & 4 \\
			U_7 & \frac{1}{6} & \frac{1}{2} & (*) & \frac{1}{2} & \frac{1}{6} & & & & & & & 4\\
			\br
		\end{array}$
	\end{table}
	
\begin{figure}
		\centering
		\includegraphics[width=0.6\linewidth]{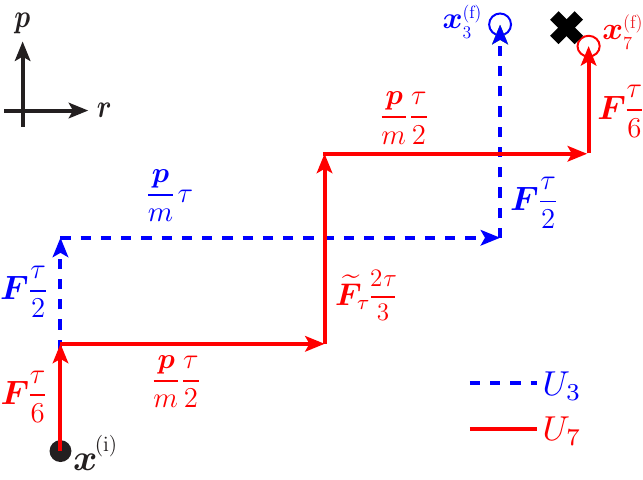}
		\caption{The ST approximations of \eref{eq:STDecomp} are leapfrog algorithms when transferred to classical dynamics in phase space according to \eref{eq:connection1} and \eref{eq:connection2}. Here, we depict the 2nd-(4th-) order leapfrog algorithm $U_3$ ($U_7$) schematically. $\bi{x}^{(\mathrm i)}$ represents the initial phase-space point, $\bi{x}_3^{(\mathrm f)}$ and $\bi{x}_7^{(\mathrm f)}$ represent the final points for $U_3$ and $U_7$, respectively, and the black cross marks the exact final phase-space point.}
		\label{fig:Leapfrog}
	\end{figure}

	Bearing in mind \eref{eq:connection1} and \eref{eq:connection2}, we can use the same approximations $U_N$ (with the same coefficients $\alpha_i$ and $\beta_i$) for classical systems. The exponential factors in \eref{eq:STDecomp} are translations in position or momentum, and are therefore symplectic transformations, such that $U_N$ induces a symplectic classical algorithm as well. For example, \fref{fig:Leapfrog} depicts the classical algorithm for $U_3$, commonly known as leapfrog algorithm, where the classical force
	\begin{equation}
		\vb{F} \left( \vb{r} \right) = - \grad V{ \left( \vb{r} \right)}
	\end{equation}
	translates momenta, while $\vb p$ translates positions. Note that $\vb r$ gets translated after $\vb p$ is propagated for only half a time step, which explains the nomenclature of the leapfrog algorithm and makes it a 2nd-order TEA.

	The TEA based on $U_7$ is a special case. It has been used in a number of studies of both classical and quantum systems \cite{Laskar+1:01, Omelyan+2:02, Skokos+1:10, Dehnen+1:17, Forbert+1:01, Chin+1:02, Chin+1:05, Lehtovaara+2:07}. The $U_7$ approximation was first discovered in \cite{Chin:97}, reconsidered in \cite{Omelyan+2:02}, but later independently obtained through an entirely different approach in \cite{Chau+3:18}. While $U_7$ is obtained in \cite{Chin:97} by removing the 3rd-order error manually, Suzuki considered the inclusion of gradient terms in the factorization of the time-evolution operator \cite{Suzuki:95}. Omelyan \textit{et~al.} in \cite{Omelyan+2:02} introduced a more general factorization scheme that includes $U_7$. The derivation in \cite{Chau+3:18} starts with the 7-factor ST approximation ${\alpha_1=\alpha_4=1/6}$, ${\alpha_2=\alpha_3=1/3\pm1/\epsilon}$, ${\beta_1=\beta_3=1/2\mp\epsilon/24}$, and ${\beta_2=-\epsilon^2/36}$. While maintaining the accuracy of ${\mathcal O(t^4)}$ and using the relation
	\begin{equation}
		\Exp{- \I f(\vb{R})} g(\vb{P}) \Exp{\I f(\vb{R}) } = g \left(\vphantom{X_{X_{X_{m}}}}\vb{P} + \hbar \grad f(\vb{R}) \right) ,
	\end{equation}	
	we reduce $U_7$ to a 5-factor approximation in the limit ${\epsilon \rightarrow 0^{+}}$, with the result that a gradient term replaces the exponential factor associated with $\alpha_2$, see~\tref{tab:STCoeff}.
	The corresponding classical algorithm with
	\begin{equation}
		\tilde{\vb{F}}_t \left( \vb{r} \right) = - \grad { \left( V - \frac{1}{48 m} \left[ t \grad V \right]^2  \right)}
	\end{equation}
	is illustrated in \fref{fig:Leapfrog}.

	\section{Benchmarking against exact dynamics} \label{sec:compare}
	
	In the following, we shall benchmark the TEA defined in \sref{sec:leapfrog} against exact dynamics of various classical and quantum systems, see \tref{tab:systems}, and identify $U_7$ as the most efficient TEA among the three 4th-order algorithms given in \tref{tab:STCoeff}. We quantify the performance of each algorithm (with the exception of the non-periodic honeycomb system) by the period errors
	\begin{equation}
		\epsilon_{\mathrm{C}} = \sqrt{ \left( \vb{x}_T - \vb{x}_0 \right)^2 + \left( \vb{p}_T - \vb{p}_0 \right)^2 }
	\end{equation}
	and\footnote{The scalar product ${\braket{\Psi_1(t_1)}{\Psi_2(t_2)}}$ denotes ${\int(\mathrm{d}\vb a)\, \psi_1(\vb a,t_1)^*\,\psi_2(\vb a,t_2)}$ for any basis ${\{\ket{\vb a}\}}$.}
	\begin{equation}
		\epsilon_{\mathrm{Q}} = \ABS{\braket{\Psi(T)}{\Psi(0)} - 1}
	\end{equation}
	for classical and quantum systems, respectively. Here, ${ \left( \vb{x}_0, \vb{p}_0 \right) }$ and ${ \left( \vb{x}_T, \vb{p}_T \right) }$ are the initial (${t=0}$) and final (${t=T}$) phase-space positions, while $\Psi(0)$ and $\Psi(T)$ are the initial and final wave functions. The large exact revival period of the Rydberg state makes $\epsilon_{\mathrm{Q}}$ difficult to compute in practice. As an alternative, we determine the overlap error
	\begin{equation}\label{eq:overlaperror}
		\epsilon_{\mathrm{O}} = \ABS{\braket{\Psi_{\mathrm{ex}}(T)}{\Psi(T)} - 1},
	\end{equation}
	where $\Psi_{\mathrm{ex}}(T)$ and $\Psi(T)$ are the exact and approximate final wave functions, respectively. 
	
	\begin{table}
		\centering
		\caption{Our test set of classical and quantum systems with known dynamics. $E$ is the total energy of the system, $\omega$ is the angular frequency of the periodic motion, $a$ is the length of the pendulum, $\theta$ is the deflection angle of the pendulum, $n$ and $\ell$ are the principal and angular momentum quantum number, $e$ is the electron charge, $\bar{n}$ is the average principal quantum number for the Rydberg wave packet, and $\K(\;)$ is the complete elliptic integral of the first kind \cite{Abramowitz+1:72}. }
		\label{tab:systems}
		\vspace{0.3cm}
		\setlength\tabcolsep{0.22em}
		\begin{indented}
			\item[]\begin{tabular}{@{}c c c c}
				\br
				& system & potential energy ${ V {\left( \vb{r} \right)} }$ & period $ T $  \\ \mr
					\multirow{3}{*}[-1.3em]{\begin{sideways}classical\end{sideways}} & pendulum & $ m a^2 \omega^2 (1 - \cos \theta) $ & $ \frac{4}{\omega} \vphantom{\Big(\Big)} \K {\left( \frac{E}{2 m \omega^2 a^2} \right)} $  \\
				& Kepler orbit & $ - \frac{1}{r} $ & $ \frac{\pi}{\sqrt{2}} \ABS{E}^{-\frac{3}{2}} $  \\[0.5em]
				& \parbox{7em}{\centering 2D honeycomb\\ potential} & $ 3 + 2 \sum\limits_{n=1}^{3} \cos{\Big(}\cos( \frac{n\pi}{3} )x + \sin ( \frac{n \pi}{3} )y {\Big)} $ & chaotic dynamics  \\
				\mr
				\multirow{3}{*}[-1em]{\begin{sideways}quantum\end{sideways}} & \parbox{7em}{\centering 2D harmonic\\ oscillator} & $ \vphantom{\Bigg(\Bigg)}\frac{1}{2} m \omega^2 \vb{r}^2 $ & $ \frac{2 \pi}{\omega}  $  \\
				& \parbox{7em}{\centering 3D Davidson\\ potential} & $ \frac{1}{2} m \omega^2 \vb{r}^2 + \frac{ \hbar^2 }{2 m r^2} $ & $ \frac{2 \pi}{\omega} \left( 2 n + 1 + \sqrt{ \left( \ell + \frac{1}{2} \right)^2 + 1} \right)^{-1} $  \\[0.8em]
				& Rydberg atom & $  -\frac{e^2}{r} $ & $ 2 \pi \bar{n}^3 \frac{\hbar^3}{\sqrt{m^3 e^7}}  $ \\
				\br
			\end{tabular}
		\end{indented}
	\end{table}	

	Having determined the figures of merit, we proceed with two ways of benchmarking. The first is to evolve the system for one period and a fixed number of steps $N$, resulting in
	\begin{equation}
		\epsilon\left(U_7\right) < \epsilon\left(U_{11}\right) < \epsilon\left(\mathrm{RK}4\right) < \epsilon\left(U_7'\right) < \epsilon\left(U_3\right)  \label{eq:order}
	\end{equation}
	for the period errors (RK4 is considered only for classical systems). Clearly, for $N$ spanning several orders of magnitude, $U_7$ is the most accurate 4th-order TEA among those considered here. Our data are consistent with the fact that the log-log graph for error vs.~$N$ has a slope of $-n$ for an $\mathcal O(t^n)$-TEA: Averaging over all systems studied, we obtain the slopes $ -1.98 \pm 0.04$ and $-4.02 \pm 0.03 $ for $U_3$ and $U_7$, respectively.

	In practice, the number of steps $N$ does not matter as much as the computation time, which serves as our second type of benchmarking. The computation time for the ST-based algorithms scales with the number of factors in the ST approximation. For example, while both $U_7$ and $U_{11}$ are 4th-order approximations, $U_7$ consists of only five factors and roughly takes half the computing power for accomplishing one time step, compared with the eleven-factor approximation $U_{11}$. This scaling is of particular importance for quantum applications, where costly Fourier transforms are invoked to switch between position and momentum space, see \eref{U3FFT}. For all cases considered, the computation time needed to achieve a fixed accuracy follows the sequence in \eref{eq:order}. This shows that $U_7$ is also the most \emph{efficient} algorithm to achieve a given accuracy. In the remaining sections we substantiate these general results by more detailed discussions of the systems defined in \tref{tab:systems}.\footnote{We set ${\hbar = m = \omega = 1 }$ during numerical simulations.}

	\section{Classical Systems} \label{sec:classical}
	
	\begin{figure}
		\centering
		\includegraphics[scale=0.6]{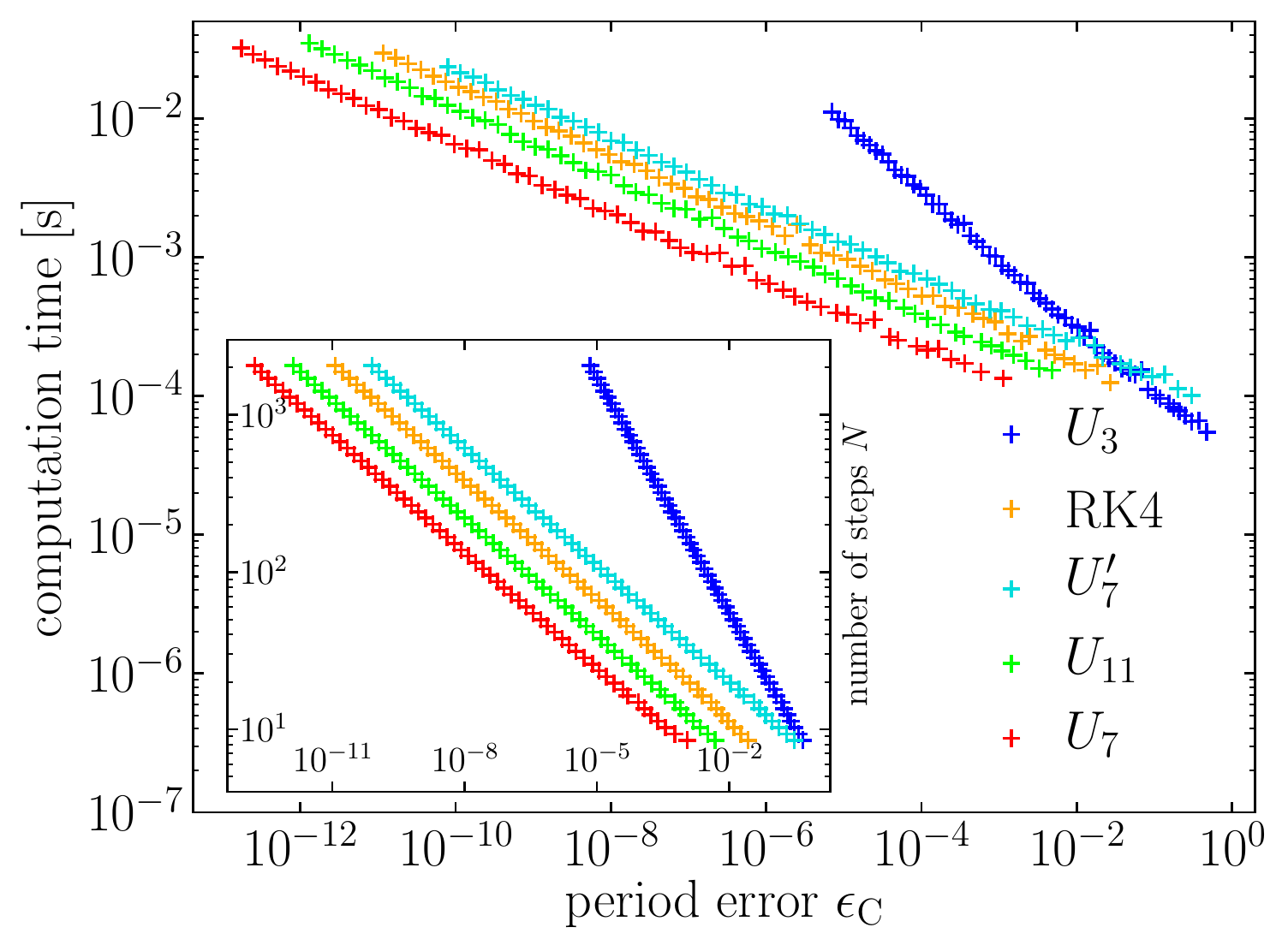}

		\caption{$U_7$ achieves the highest accuracy in the case of a classical pendulum, both in terms of computation time (main plot) and number of steps over one period (inset), outperforming in particular its direct competitor $U_{11}$ \cite{Hatano+1:05} and the 4th-order Runge--Kutta method.}%
		\label{fig:pendulum}
	\end{figure}

	We begin with the textbook example of a pendulum moving in one dimension: \Fref{fig:pendulum} shows computation time (main plot) and period error $\epsilon_{\mathrm{C}}$ (inset). Both measures follow \eref{eq:order} --- an outcome we also found for the other systems listed in \tref{tab:systems}.

	\begin{figure}
		\centering
		\begin{minipage}{0.45\linewidth}
			\begin{picture}(90,170)
				\put(0,0){\includegraphics[scale = 0.5]{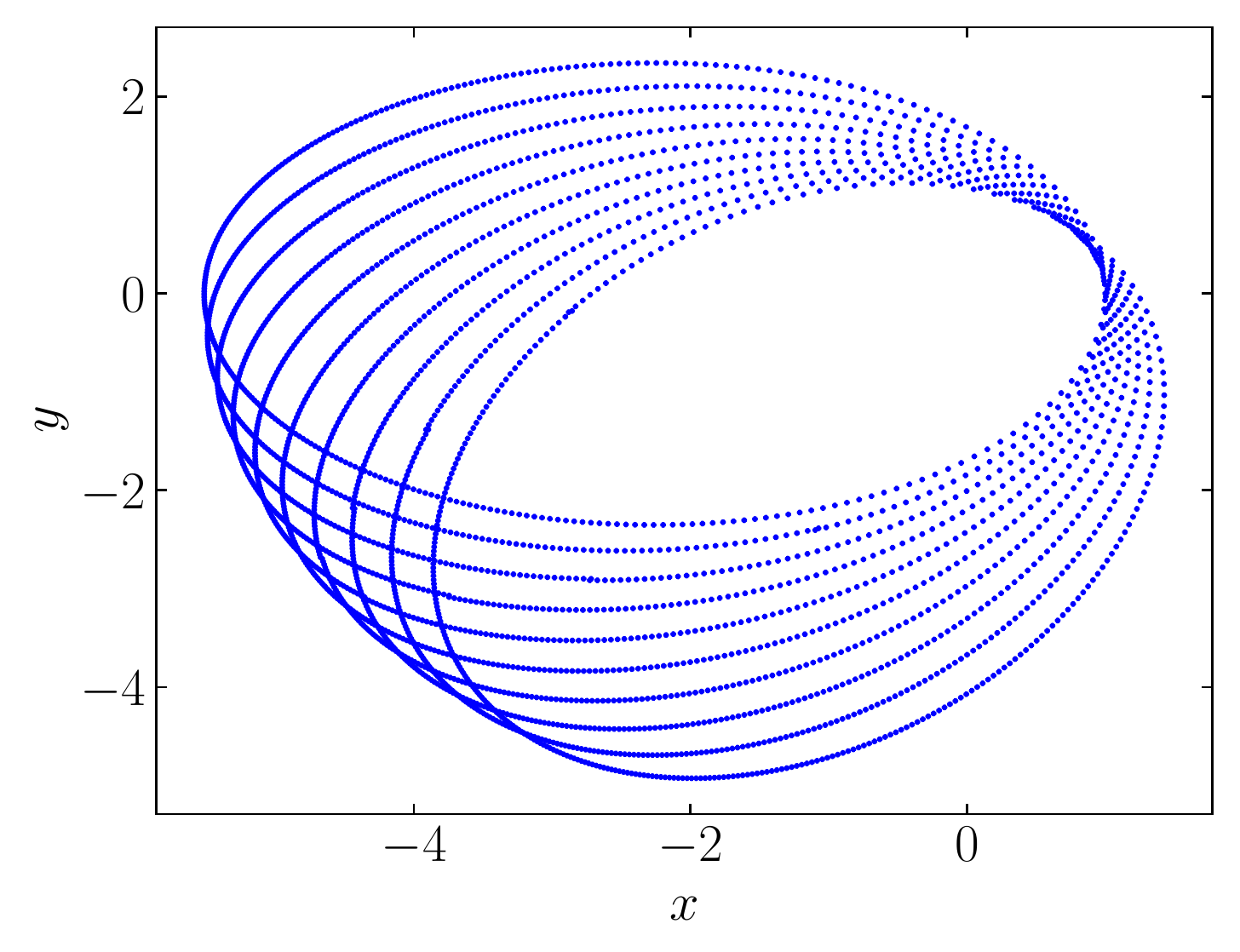}}		
				\put(30,145){(a)}	
			\end{picture}
		\end{minipage}
		~
		\begin{minipage}{0.45\linewidth}
			\begin{picture}(90,170)
				\put(0,0){\includegraphics[scale = 0.5]{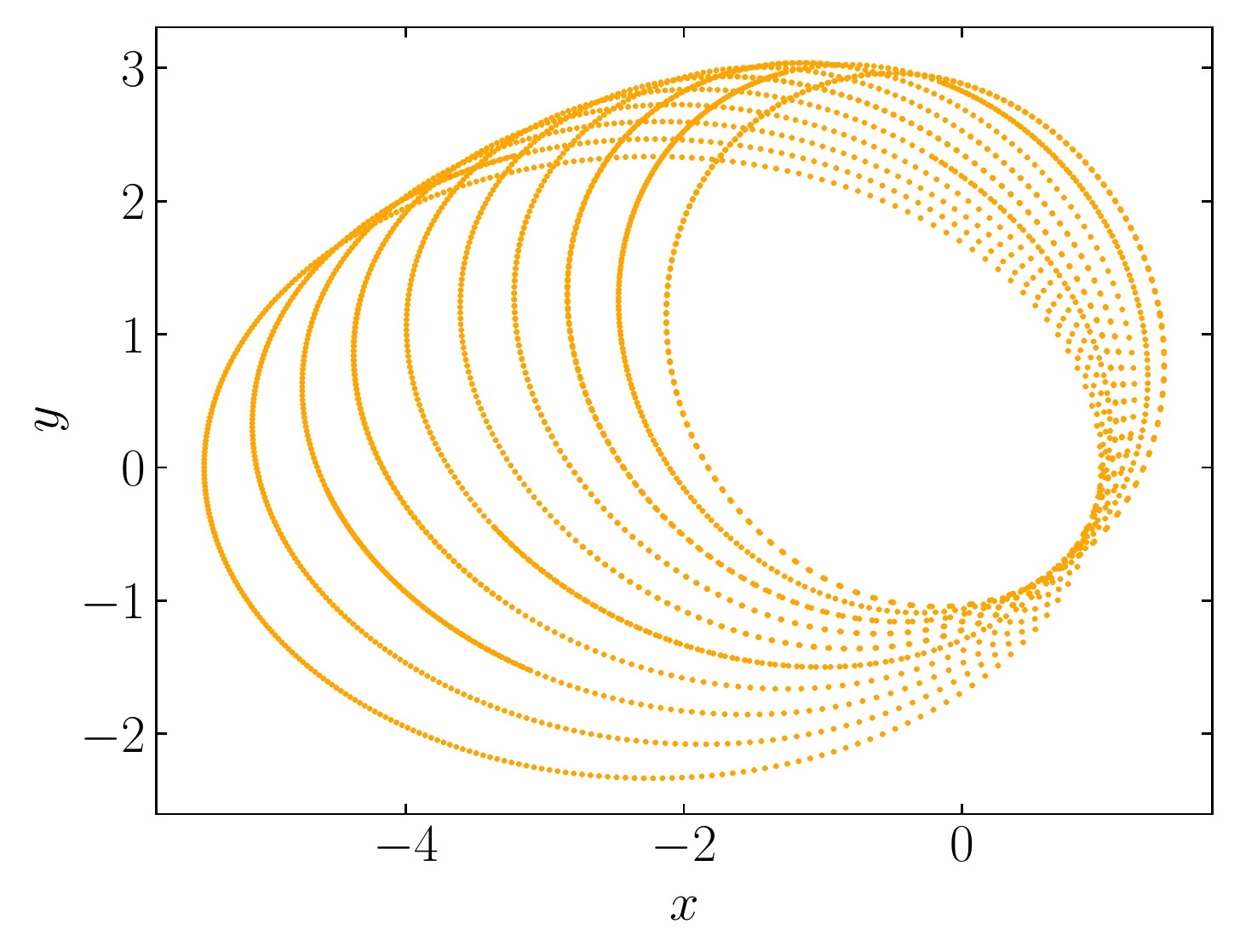}}		
				\put(30,145){(b)}	
			\end{picture}
		\end{minipage}\\  
		\begin{minipage}{0.45\linewidth}
			\begin{picture}(90,170)
				\put(0,0){\includegraphics[scale = 0.5]{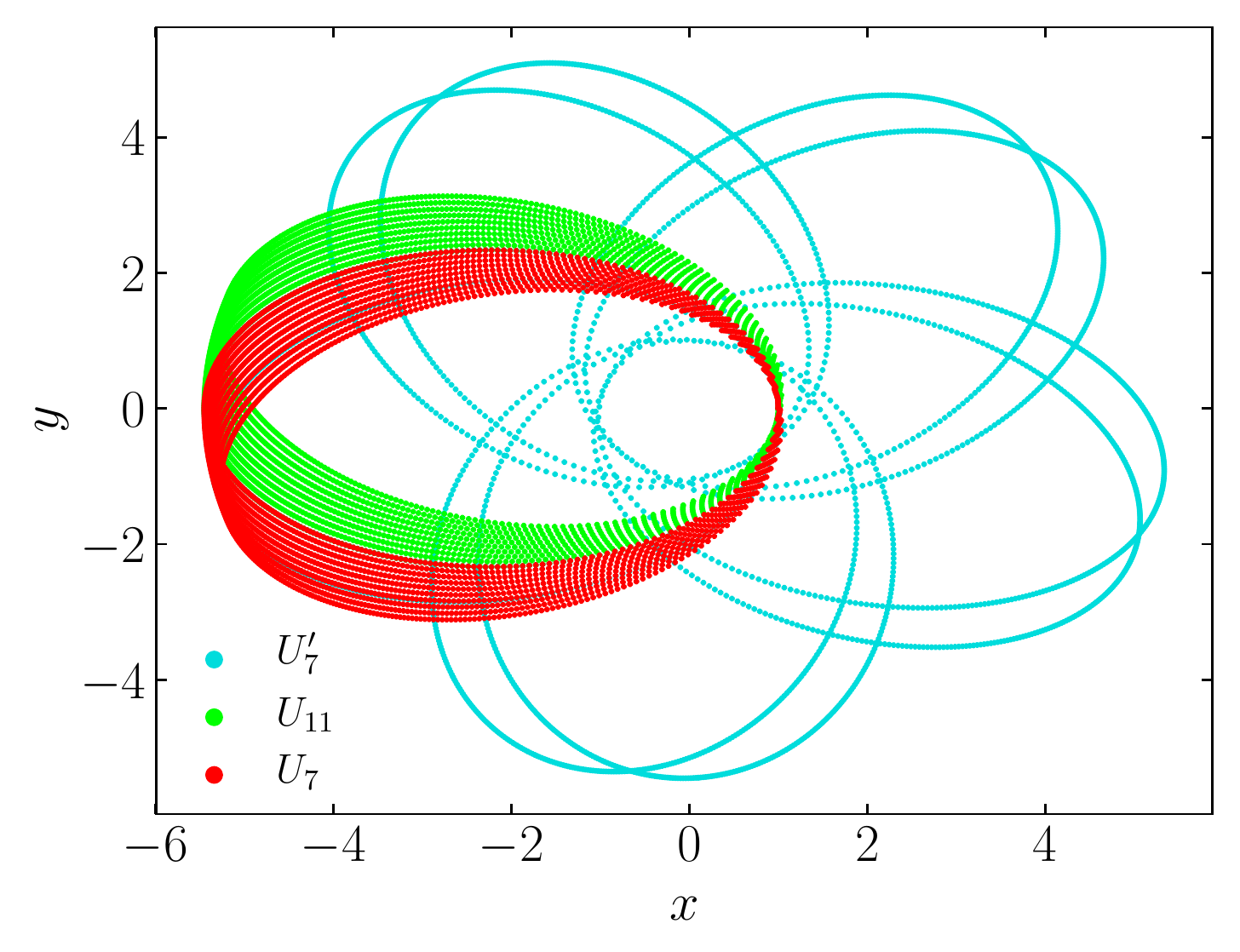}}		
				\put(30,145){(c)}	
			\end{picture}
		\end{minipage}
		~
		\begin{minipage}{0.45\linewidth}
			\begin{picture}(90,170)
			\put(0,0){\includegraphics[scale = 0.5]{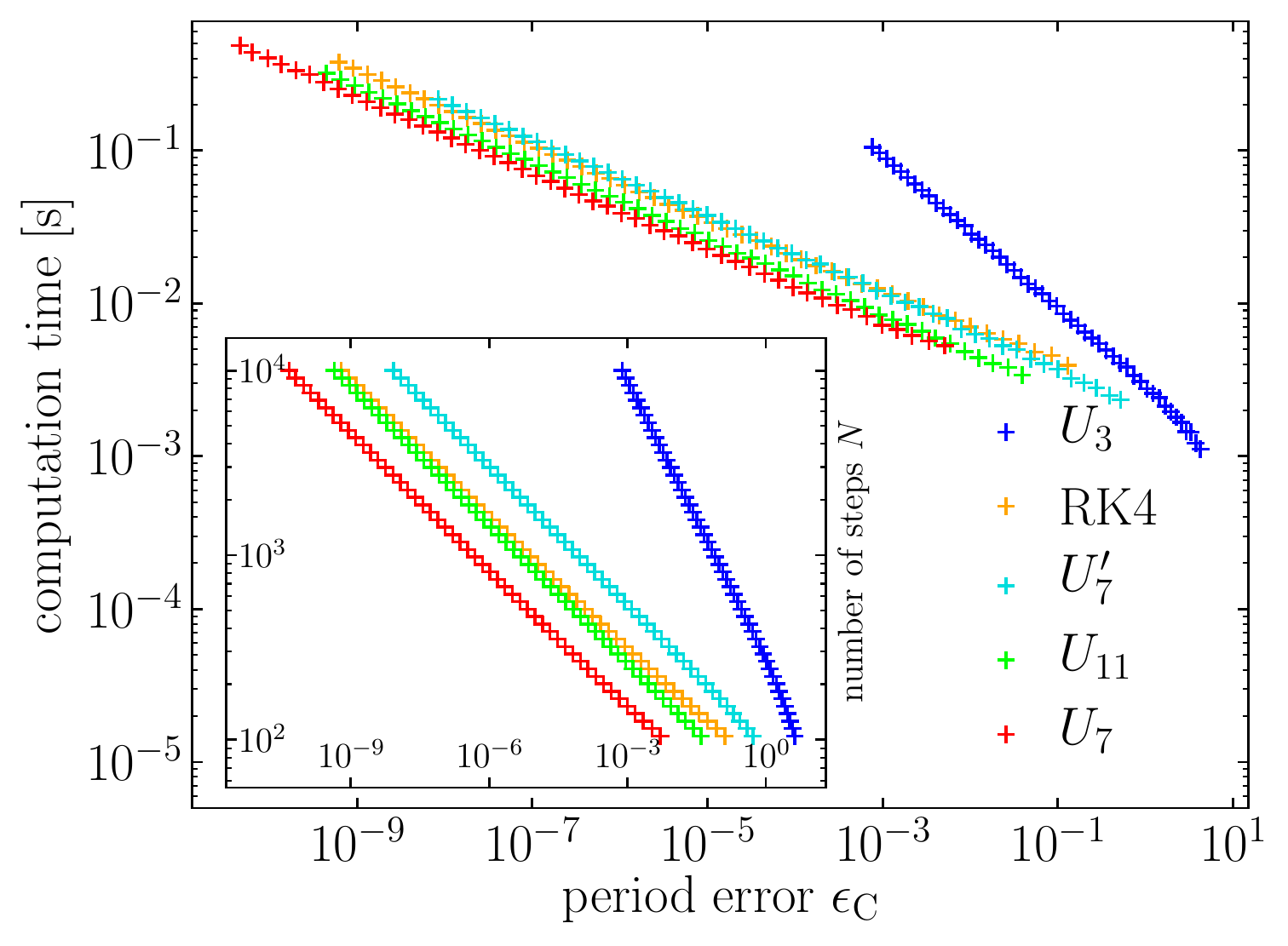}}		
			\put(35, 135){(d)}	
			\end{picture}
		\end{minipage}

		\caption{Position snapshots of a point particle moving in a central gravitational field. A precession of the Kepler orbits becomes visible when propagating for (a) 200 periods with $U_3$, (b) 64000 periods with RK4, and (c) 64000 periods with $U_7$, $U_7'$, and $U_{11}$. Clearly, $U_7$ and $U_{11}$ are the most stable TEA here. Note that RK4 does not preserve the area enclosed by the orbit, in contrast to the symplectic TEA based on ST factorizations. Panel (d) demonstrates that $U_7$ outperforms the other 4th-order algorithms in both computation time (main plot) and number of time steps per period (inset), see~\eref{eq:order}.}
		\label{fig:kepler}
	\end{figure}

	Next, we consider the two-dimensional Kepler planetary system, which harbors a potentially troublesome singular potential energy, and test the performance of our classical TEA by observing orbits in position space. Numerical algorithms for predicting trajectories in real-world gravitational fields are sought-after tools for predicting flight paths of satellites and spacecrafts in astronomy and astrophysics \cite{Kinoshita+2:90,Bravetti+3:20}. For our benchmarking exercise, however, we work with the textbook Kepler problem of a point particle in the field of another point particle with infinite mass. In that case the exact trajectories are stable ellipses without precession. Our least accurate algorithm $U_3$ produces a substantial precession over 200 periods due to the numerical errors beyond second order, see \fref{fig:kepler}(a). However, its symplectic nature preserves the area of the orbits. In contrast, the 4th-order algorithm RK4 is not symplectic, such that the area enclosed by the orbits decays as shown in \fref{fig:kepler}(b). All orbits in \fref{fig:kepler}(c) are area-preserving as expected, but both $U_7$ and $U_{11}$ are evidently superior to $U_7'$. The competition between $U_7$ and $U_{11}$ is settled in \fref{fig:kepler}(d) where the computation time and period error $\epsilon_\mathrm{C}$ both follow \eref{eq:order}. 

	\begin{figure}
		\centering 
		\begin{minipage}{0.33\linewidth}
			\begin{picture}(100, 150)
				\put(0,0){\includegraphics[scale = 0.5]{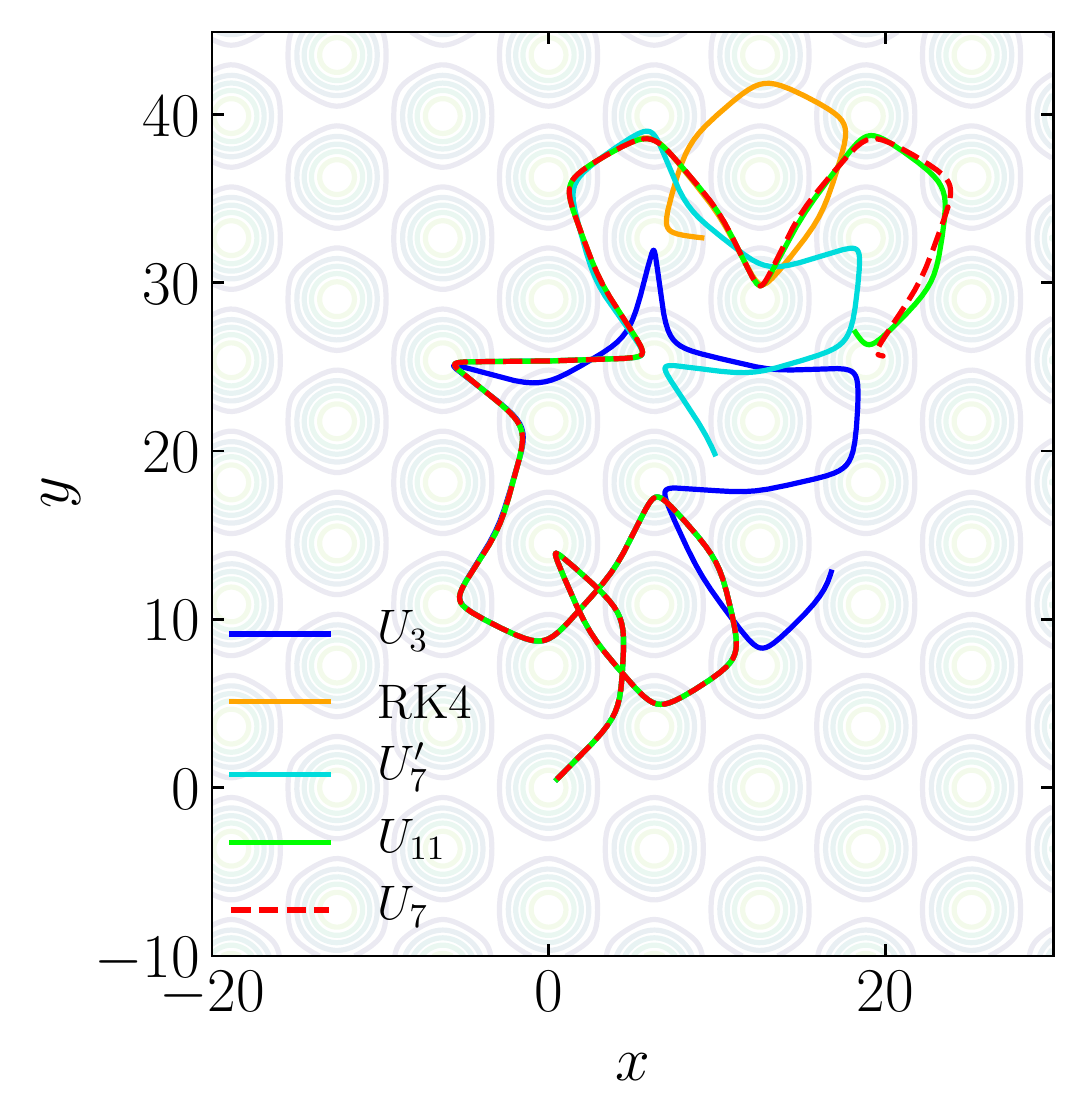}}
				\put(35, 145){(a)}
			\end{picture}
		\end{minipage}~
		\begin{minipage}{0.33\linewidth}
			\begin{picture}(100, 150)
			\put(0,0){\includegraphics[scale = 0.5]{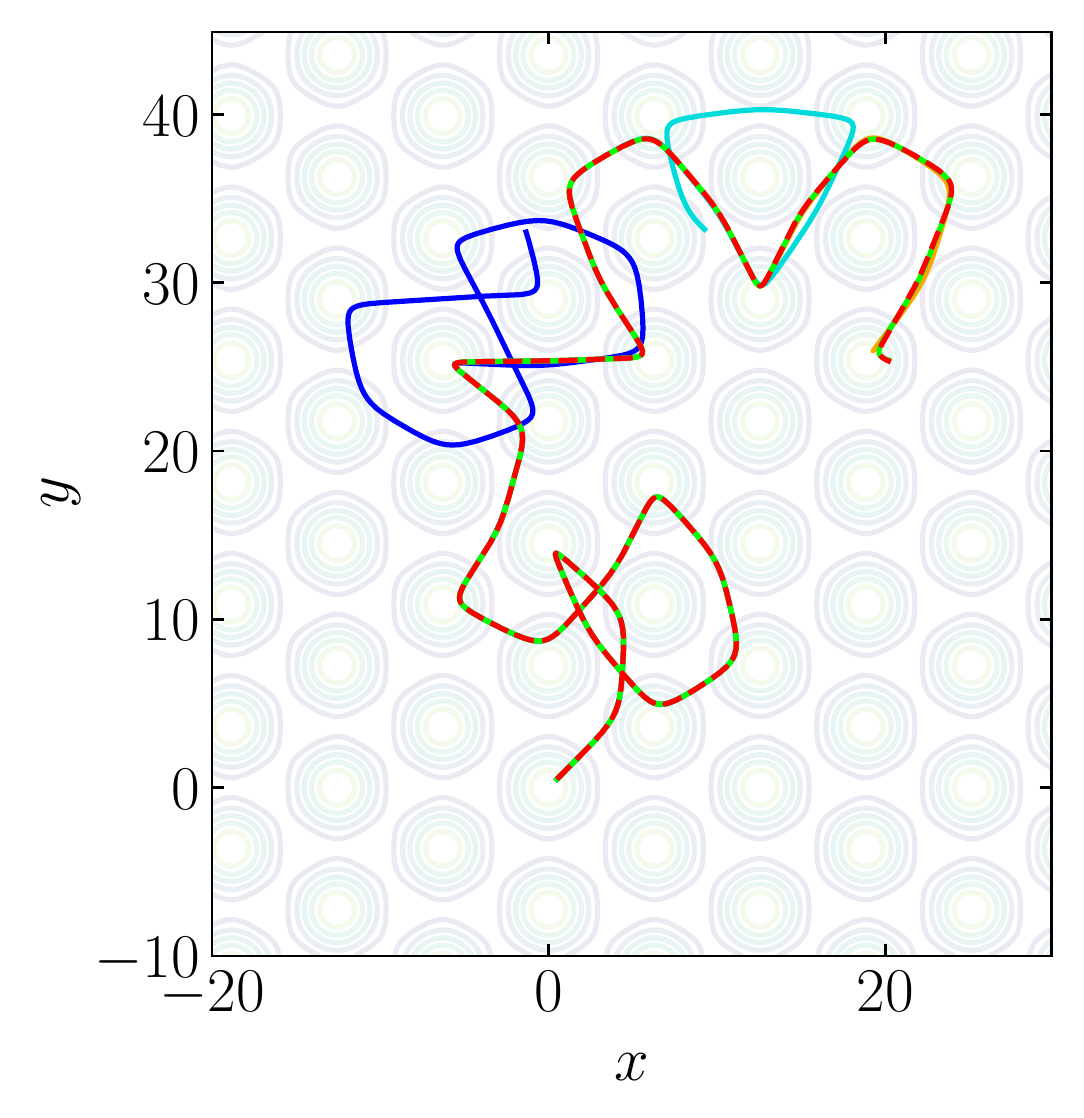}}
			\put(35, 145){(b)}
			\end{picture}
		\end{minipage}~
		\begin{minipage}{0.33\linewidth}
			\begin{picture}(100, 150)
			\put(0,0){\includegraphics[scale = 0.5]{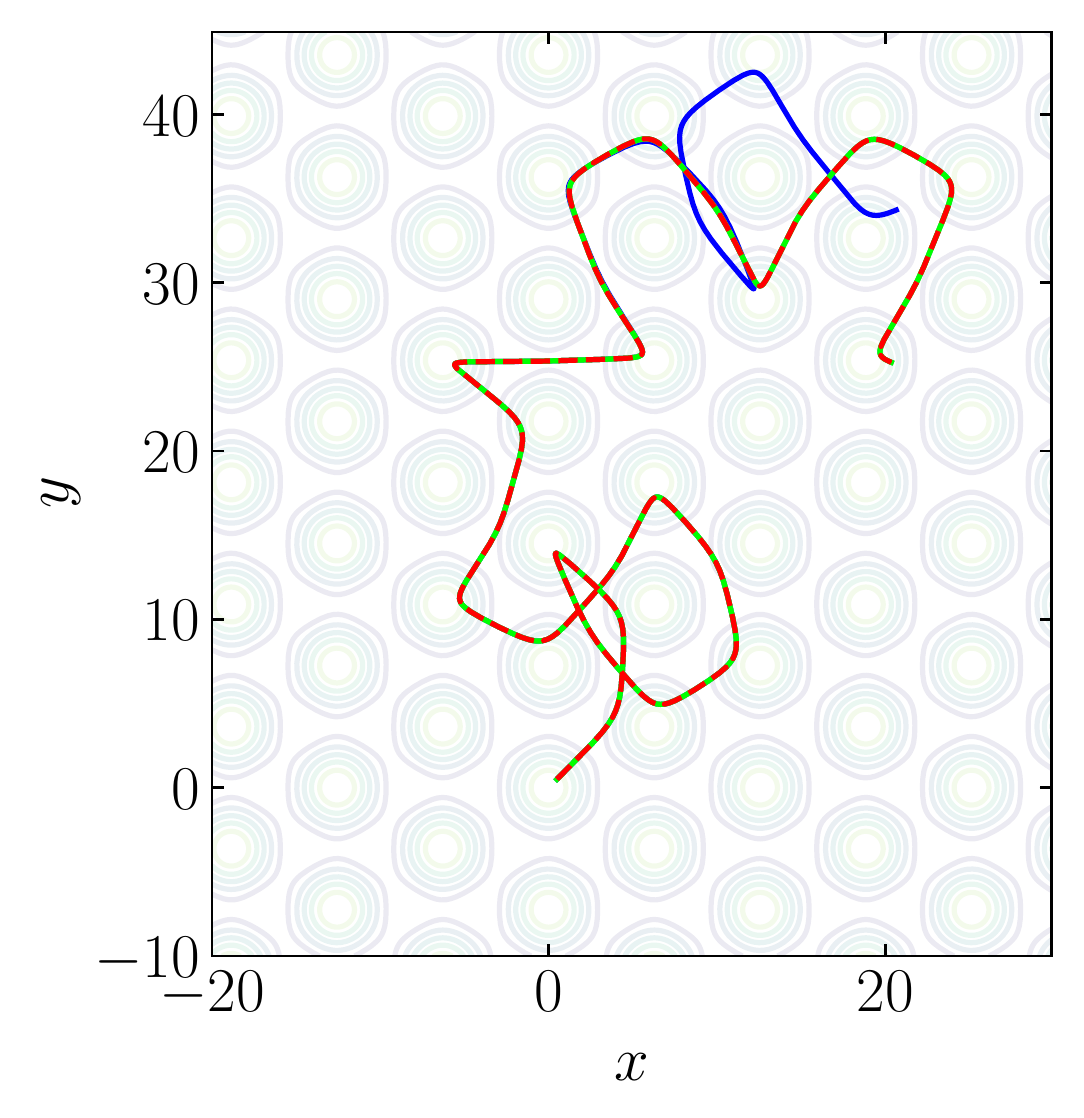}}
			\put(35, 145){(c)}
			\end{picture}

		\end{minipage}
		\caption{The trajectory of a classical particle in a honeycomb potential is most efficiently predicted by the $ U_7 $ TEA since it converges to the true trajectory using a time step of (a) ${ \dd{t} = 0.02 \, \mathrm{s} }$, while the other 4th-order algorithms converge only for smaller time steps (b) ${ \dd{t} = 0.01 \, \mathrm{s} }$ and (c) ${ \dd{t} = 0.001 \, \mathrm{s} }$.}
		\label{fig:honey}
	\end{figure}

	Higher-order TEA are also relevant for deterministically chaotic systems, for which the predicted time evolution is sensitive to minute numerical errors in the algorithm itself and/or the initial state. For our performance test, we choose a classical particle in a honeycomb potential \cite{Porter+2:16}, see \tref{tab:systems}, and benchmark the various algorithms by the time step required for convergence to the true trajectory (which is the trajectory obtained for infinitesimal time steps). The results shown in \fref{fig:honey} are consistent with the order in \eref{eq:order}. Overall, we find that $U_7$ is the best performing 4th-order TEA for classical systems.

	\section{Quantum Dynamics} \label{sec:quantum}
	
\begin{figure}
		\centering
		\begin{picture}(200,200)
			\put(0,0){\includegraphics[scale=0.6]{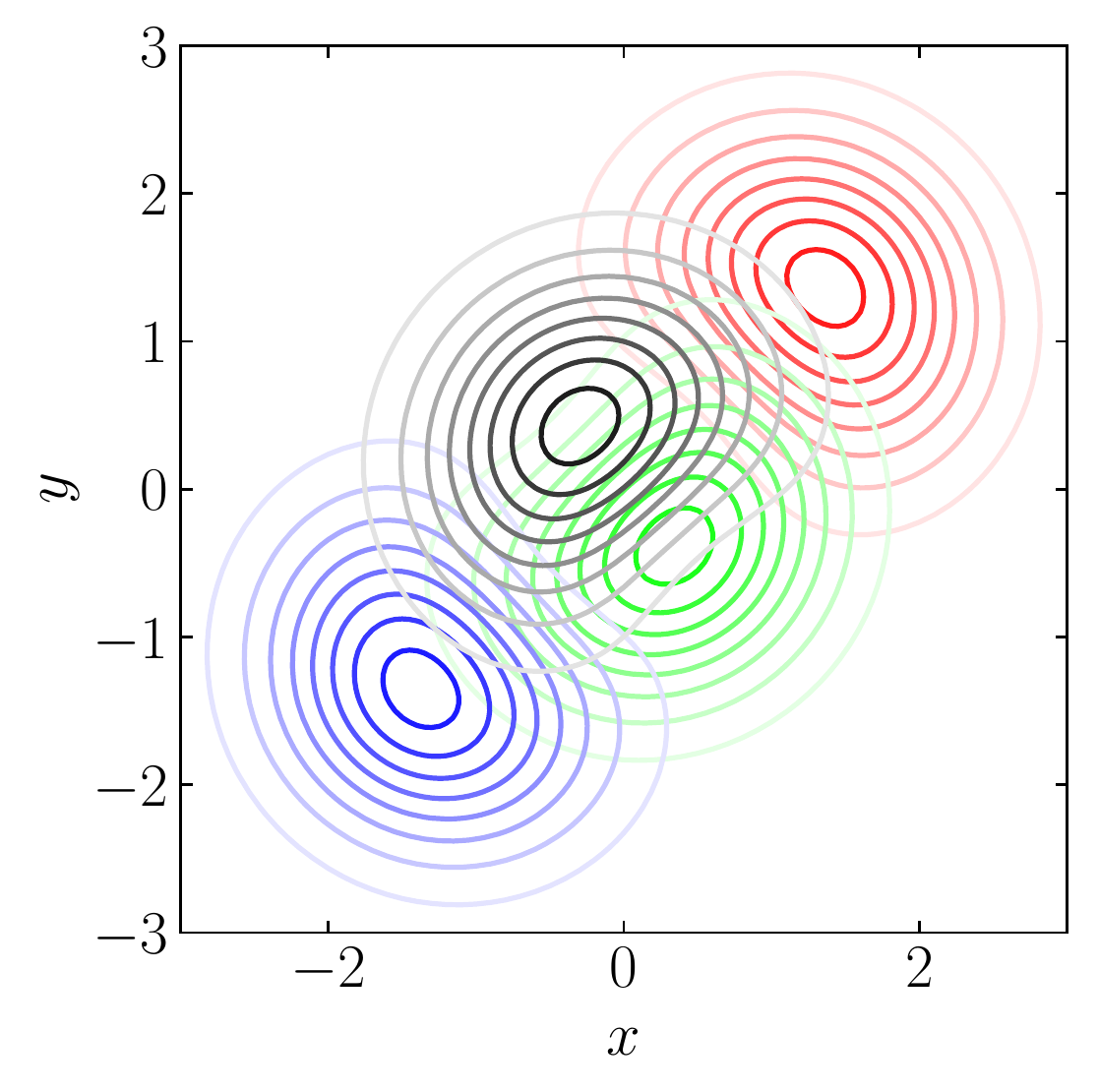}}
			\put(200,100){\begin{tabular}{c c}
					\br
					TEA & $\epsilon_{\mathrm{Q}}$ \\
					\mr
					$U_3$ & $3 \times 10^{-2}$ \\
					$U_7'$ & $ 2 \times 10^{-5}$ \\
					$U_{11}$ & $ 3 \times 10^{-7}$ \\
					$U_7$ & $7 \times 10^{-8}$ \\
					\br
			\end{tabular}}
		\end{picture}
		
		\caption{Density contour lines (larger opacity for larger densities) of a rotating wave packet in a 2D harmonic oscillator trap over an exact period $T$ divided into ${N=100}$ time steps. The snapshots are spaced at ${t=0}$ (red), $ \frac{1}{4}T $ (green), $ \frac{1}{2}T $ (blue), and $ \frac{3}{4} T$ (black). $U_7$ exhibits the smallest period error. }
		\label{fig:2DHO}
	\end{figure}

	In this section, we shall benchmark three quantum systems and start with the textbook example of the harmonic oscillator in two dimensions, see \tref{tab:systems}. As initial wave function we choose the Gaussian wave packet (see, for example, \cite{Cohen-Tannoudji+2:91})
	\begin{equation}
		\psi(\vb r,0) = \frac{x - \I y}{\sqrt{3 \pi}} \Exp{- \frac12\left((x-1)^2 + (y-1)^2\right) }
	\end{equation}
	with nonzero angular momentum resulting in the evolution along an ellipse as shown in \fref{fig:2DHO}. For all practical purposes, our performance test on period error $\epsilon_{\mathrm{Q}}$ and computation time replicates the results shown in figures~\ref{fig:pendulum} and \ref{fig:kepler}(d). \pagebreak\Fref{fig:2DHO} illustrates one test run, which divides one period into ${N=100}$ time steps. The quality of the TEA follows the sequence in \eref{eq:order} as expected from the classical dynamics benchmarking.

	Our next example is the singular Davidson potential \cite{Davidson:32} in three dimensions, see \tref{tab:systems}, which is used, for instance, to calculate rotation-vibrational spectra of diatomic molecules \cite{Rowe+1:98}. The eigenstates $\ket{n,\ell,m_\ell}$ of the Hamiltonian can be found in \cite{Davidson:32}. In our simulation we choose ${(n, \ell, m_\ell) = (1, 20, 20)}$ for the quantum numbers of the initial state, whose low probability amplitudes around the origin are then numerically more tractable when propagated with a potential-dependent exponential factor. \pagebreak Again, we find $U_7$ to outperform the other TEA in both overlap error and computation time, with \eref{eq:order} obeyed. Evolving one period in ${N=100}$ steps, we find 
	\begin{equation}
		\begin{tabular}{c c c c c}
			\br
			TEA & $U_3$  & $U_7'$ & $U_{11}$ & $U_7$ \\
			\mr
			$\epsilon_\mathrm{Q}$ & $2 \times 10^{-6}$ & $ 2 \times 10^{-11} $ & $ 6 \times 10^{-12} $ & $ 2 \times 10^{-12}$ \\
			\br
		\end{tabular} ,
	\end{equation}
	consistent with \eref{eq:order}.

\begin{figure}
		\centering
		\begin{minipage}{0.33\linewidth}
			\hspace{1.2cm}
			\begin{picture}(150,160)
			\put(-50,2){\includegraphics[scale = 0.5]{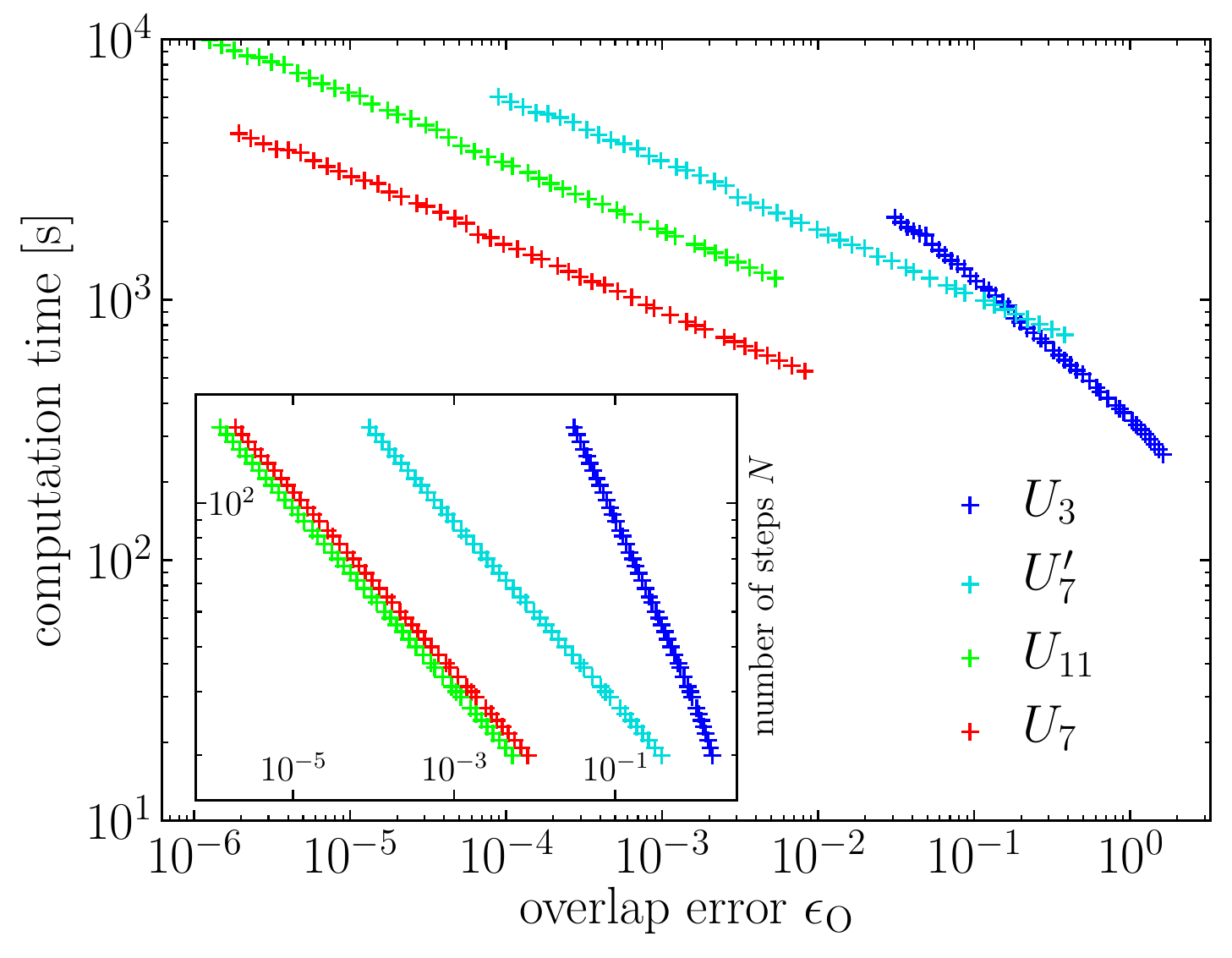}}		
			\put(140,145){\footnotesize(a)}	
			\end{picture}
		\end{minipage}\\[0.5cm]
		\begin{minipage}{0.66\linewidth}
			\begin{picture}(210,160)
			\put(10,0){\includegraphics[width=12.2cm]{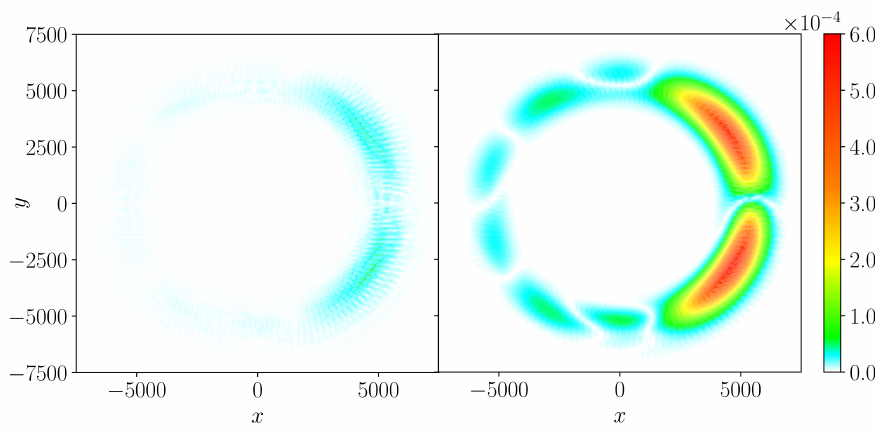}}		
			\put(165,145){\footnotesize(b)}	
			\put(310,145){\footnotesize(c)}	
			\end{picture}
		\end{minipage}~
		\caption{While panel (a) reveals $ U_{11} $ as the slightly better TEA compared with $ U_7 $ in terms of the overlap error for fixed number of time steps, $U_7$ still has a significant advantage over $ U_{11} $ in terms of computation time (propagation for 1000 seconds): Panel (b) shows the relative difference ${\left. \ABS{\ABS{\psi_{U_7}(\vb r)}^2-\ABS{\vphantom{\psi_{U_7}}\psi_{\mathrm{ex}}(\vb r)}^2}\right/\max{\ABS{\psi_{\mathrm{ex}}(\vb r)}^2}  }$ between the exact solution and $ U_7 $, whose performance is an order of magnitude better than that of $ U_{11} $, which is depicted in panel (c).}
		\label{fig:3DRyd}
	\end{figure}

	\pagebreak Rydberg wave packets present a more advanced test for our TEA. The high energy electron in a Rydberg atom is treated as the Gaussian wave packet\footnote{Here, $\bar{n}$ and $\sigma_n^2$ are the mean and the standard deviation of the Gaussian wave packet, respectively.}
	\begin{equation}
		\Psi(t) = \frac{1}{(2 \pi \sigma_n^2)^{1/4}} \sum_{n=1}^{\infty} \Exp{ -\frac{(n-\bar{n})^2}{4 \sigma_n^2}} \, \Exp{ \frac{\I t}{2 n^2} }\,\Psi_{n,n-1,n-1}(0),
		\label{eq:exactRyd}
	\end{equation}
	which represents a superposition of basis states $\ket{\psi_{n,\ell, m_\ell}}$ with quantum numbers ${(n, \ell, m_\ell) = (n, n-1, n-1)}$, see, for example, \cite{Gaeta+1:90}. For our numerical simulation we choose ${\bar{n} = 75}$, ${\sigma_n = 2.5}$ to obtain a narrow Gaussian wave packet, with $n$ only ranging from $73$ to $77$ for computational simplicity. Since the periods for the individual eigenstates follow straightforwardly from their energies, we can compute the exact revival time $T_R$ for such a superposition state. However, $T_R$ is too large in view of our computing resources, even for the modest number of eigenstates employed here. As an alternative, we therefore propagate the initial state by one Kepler orbit (classical period $2 \pi \bar{n}^3$) and calculate the overlap error $\varepsilon_{\mathrm{O}}$ from \eref{eq:overlaperror} using the exact time-evolved state in \eref{eq:exactRyd}. In contrast to all our previous examples, $U_{11}$ delivers a slightly better result than $U_7$ in terms of $\varepsilon_{\mathrm{O}}$, see \fref{fig:3DRyd}(a). But $U_7$ still outperforms $U_{11}$ by an order of magnitude in terms of the more relevant computation time (here, set to $1000$ seconds). This is demonstrated in the panels (b) and (c) of \fref{fig:3DRyd}, which show the relative difference between the exact solution and the final wave functions based on the $U_7$ and $U_{11}$ algorithms, respectively.

	\section{Conclusion} \label{sec:conclusion}
	
	In our study, we examine the performance of time-evolution algorithms derived from Suzuki--Trotter factorizations of the time-evolution operator. This split-operator approach reveals the link between classical and quantum dynamics and permits a straightforward numerical implementation. When benchmarking against exact data for classical and quantum systems, we find that the fourth-order algorithm $U_7$, for which a new concise derivation has been found recently \cite{Chau+3:18}, outperforms established algorithms (including the popular fourth-order Runge--Kutta method) in both accuracy and computation time. These results are consistent with the findings reported in \cite{Laskar+1:01, Omelyan+2:02, Skokos+1:10, Dehnen+1:17, Forbert+1:01, Chin+1:02, Chin+1:05, Lehtovaara+2:07}. $U_7$ provides a valuable approximation beyond the textbook systems studied here and provides immediate practical benefits for addressing time-evolution problems at the forefront of research. In addition, our pedagogical presentation builds on concepts that are familiar to undergraduate students in physics and engineering. We therefore hope that this article, together with the accompanying user-friendly numerical software, will contribute to bridging the often encountered gap between the demands of today's scientific environment and the numerical skills of students in institutes of higher education.

	\section*{Acknowledgments}
	JHH acknowledges the financial support of the Graduate School for Integrative Science \& Enginnering at the National University of Singapore. This work is funded by the Singapore Ministry of Education and the National Research Foundation of Singapore.

	\newpage

	\section*{References}


\end{document}